# Using mathematical modeling to ask meaningful biological questions through combination of bifurcation analysis and population heterogeneity

Irina Kareva[1,2]

[1] EMD Serono Research and Development Institute, Merck KGaA, Billerica, MA 02370, USA

[2] Mathematical and Computational Sciences Center, School of Human Evolution and Social Change, Arizona State University, Tempe, AZ 85287, USA

ikareva@asu.edu

**Abstract**

Classical approaches to analyzing dynamical systems, including bifurcation analysis, can provide invaluable insights into underlying structure of a mathematical model, and the spectrum of all possible dynamical behaviors. However, these models frequently fail to take into account population heterogeneity, which, while critically important to understanding and predicting the behavior of any evolving system, is a common simplification that is made in analysis of many mathematical models of ecological systems. Attempts to include population heterogeneity frequently result in expanding system dimensionality, effectively preventing qualitative analysis. Reduction Theorem, or Hidden keystone variable (HKV) method, allows incorporating population heterogeneity while still permitting the use of previously existing classical bifurcation analysis. A combination of these methods allows visualization of evolutionary trajectories and making meaningful predictions about dynamics over time of evolving populations. Here, we discuss three examples of combination of these methods to augment understanding of evolving ecological systems. We demonstrate what new meaningful questions can be asked through this approach, and propose that the large existing literature of fully analyzed models can reveal new and meaningful dynamical behaviors with the application of the HKV-method, if the right questions are asked.

## Introduction

Heterogeneity is a major driving force behind the dynamics of evolving systems. When it is heritable and when it affects fitness, heterogeneity is what makes evolution possible (1–4). This comes from the fact that the environment in which the individuals interact is composed not only of the outside world (such as the resources necessary for survival, or members of other species) but also of individuals themselves. Therefore, selective pressures that are imposed on them come both from the environment and from each other. Furthermore, selective pressures that individuals experience from other members of the same population will be imposed and perceived differently depending in population composition, which in turn may be changing as a result of these selective pressures.

In a vast majority of conceptual, and often even in descriptive mathematical models of population dynamics, whether it be models of predator-prey interactions, spread of infectious diseases or tumor growth, population homogeneity is the first simplification that is made. It is not treated as homogeneity *per se*; rather, one assumes that an average rate of growth or death or infectiousness is a reasonable enough approximation if the system has already reached some kind of stabilized state of evolutionary development. However, by ignoring population heterogeneity in such a way, one ends up either ignoring natural selection or assuming that it has already "done its work". This assumption is often incorrect within the context of such models, since natural selection may be in fact a key driver behind dynamics of most systems that are of interest and importance.

Equation-based models are usually avoided in questions that require modeling high levels of heterogeneity. This is a result of the inevitable increase of system dimensionality that often accompanies attempts to account for population heterogeneity, to the point at which obtaining any kind of qualitative understanding of the system becomes nearly impossible. Assuming population homogeneity, while making systems of equations computationally and sometimes even analytically manageable, causes the loss of many aspects of system dynamics that come from intra-species interactions and natural selection.

Parametrically homogeneous systems can nevertheless still provide exceptionally valuable information about the structure of the system, which can be obtained through extensively developed analytical techniques, such as bifurcation analysis (5). A skillfully constructed bifurcation diagram can both reveal various possible dynamical regimes that a system can exhibit as a result of variations in parameter values and initial conditions, and provide analytical boundaries as functions of system parameters. This information can then be used to construct a theoretical framework for understanding a biological system that could never have been obtained experimentally.

Reduction theorem, also known as parameter distribution technique, or hidden keystone variable (HKV) method, is a method that allows building on insights obtained from bifurcation analysis while incorporating population heterogeneity (6–9). It allows investigating more fully the dynamics of an evolving system, while overcoming this problem of immense system dimensionality in a wide class of mathematical models.

This approach of course makes sense only when there exists a meaningful research question that a parametrically heterogeneous model can help answer (otherwise this becomes little more than a mathematical exercise).

In what follows, we will first briefly describe this approach and the assumptions and limitations that are associated with implementation of the HKV method. We will then describe several examples that reveal how a combination of the classical bifurcation analysis techniques with the HKV method can reveal previously inaccessible dynamical behaviors. We will conclude with a brief discussion on the possibilities for rich dynamical behaviors that still remain to be revealed in the already existing body of literature composed of fully analyzed mathematical models.

## General strategy

Assume the population of individuals is composed of clones $x_a$, and that each individual clone $x_a$ is characterized by parameter value $a$, which corresponds to a measure of some intrinsic heritable trait, such as birth rate, death rate, resource consumption rate, etc. The total population size is given by $N(t) = \sum_a x_a(t)$ if the system is discrete, and $N(t) = \int_a x_a(t)da$ is the system is continuous. Then, since different clones can grow and die at different rates, the distribution of clones within the population $P_a(t) = \dfrac{x_a(t)}{N(t)}$ can change over time due to system dynamics. Consequently, the mean value of the parameter $E^t[a]$, which now becomes a function of time, changes over time as well.

Analysis of a parametrically heterogeneous system involves the following steps:

1. Analyze autonomous parametrically homogeneous system to the extent possible using well-developed analytical tools, such as bifurcation analysis.

2. Replace parameter a with its mean value $E^t[a]$, which is a function of time.

3. Introduce an auxiliary system of differential equations, which define keystone variables that actually determine the dynamics of the system. (Note: the term "keystone" is chosen here in parallel to the notion of keystone species in ecology. Just like keystone species have disproportionately large effect on their environment relative to their abundance, keystone variables determine the direction in which the system will evolve, without being explicitly present in the original system).

4. Express the mean and variance of the distributed parameter, which now changes over time due to system dynamics, through keystone variables. The mean of the parameter can now "travel" through the different domains of the phase-parameter portrait of the original parametrically homogeneous system.

5. Calculate numerical solutions.

Exact formulation of the Reduction theorem and the theory underlying the method can be found in (7–9). A summary definitions and associated notation are provided in Table 1.

**Table 1.** Definitions and notation used in the application of the HKV method.

| Definition | Notation and explanation |
|---|---|
| **Selection system** | A mathematical model of an inhomogeneous population, in which every individual is characterized by a vector-parameter $a = (a_1, ..., a_n)$ that takes on values from set $\mathbb{A}$. |
| **Clone** $x_a$ | Set of all individuals that are characterized by a fixed value of parameter $a$. |
| **Total population size** $N(t)$ | $N(t) = \sum_{\mathbb{A}} x_a$ if the number of possible values of $a$ is finite and $N(t) = \int_{\mathbb{A}} x_a(t) da$ if it is infinite. |
| **Growth rate of a clone** $x_a$ | $\dfrac{dx_a(t)}{dt}$ |
| **Fitness of an individual within the population** | $\dfrac{dx_a(t)}{dt} / x_a(t)$ |
| **Distribution of clones within the population** | $P_a(t) = \dfrac{x_a(t)}{N(t)}$ |
| **Expected value of a distributed parameter** | For all expressions of the type $\dfrac{\int_{\mathbb{A}} f(a) x_a da}{N(t)}$, standard notation $E^t[f]$ of the expected value is used. |

## Advantages and drawbacks of the Reduction theorem

One of the most important properties of this method is that it allows reducing an otherwise infinitely dimensional system to low dimensionality.

However, as with any method, there are drawbacks to the application of the Reduction theorem. Most importantly, the transformation can be done (with some generalizations) only to Lotka-Volterra type equations of the form $x(t)' = x(t)F(t, f(E^t[a]))$, where $x(t)$ is a vector, $a$ is a parameter or a vector of parameters that characterize individual heterogeneity within the population, and where the form of $f(E^t[a])$ is system-specific. It can also increase the dimensionality of the original parametrically homogeneous system at a possible cost of auxiliary keystone equations (although these would typically still be on the order of only one or two extra equations, depending on the original system). Finally, the resulting system is typically non-autonomous, so one cannot perform standard bifurcation analysis and has to resort to calculating numerical solutions.

When studying numerical solutions of such parametrically heterogeneous systems, one can observe trajectories that one could not previously have seen in parametrically homogeneous systems. This phenomenon results from the expected value of the parameter "traveling" through the phase parameter portrait, undergoing qualitative phase transitions as it crosses the bifurcation boundaries. Furthermore, now, if there exists a complete bifurcation diagram for the specific parametrically homogeneous model, one can identify what boundaries have been crossed during system evolution.

One can also not only track the distribution of different clones within the population as the system evolves but also observe how different initial distributions of clones in the population can lead to different trajectories. One can therefore capture effect of sensitivity to initial population composition both to changes in intrinsic properties of the individuals (such as birth or death rates) or to changes in the external factors (environment) without observing chaotic behavior. This results from the fact that different clones have different fitness depending on initial population composition, since the selective pressures that are imposed on them result not only from the external environment but from surrounding clones as well.

Therefore, the HKV method allows for equation-based models to generate complex behaviors by incorporating all the properties of a complex system (2) without significantly increasing system dimensionality. Notably, unlike agent-based models, which are the standard computation tool for studying complex systems, the HKV method does not allow incorporating spatial heterogeneity.

Next, we will describe a series of examples, when application of the two methods coupled with a meaningful research question allowed answering questions and visualizing previously unobserved evolutionary trajectories.

## Example 1. Sustainability: Using a parametrically heterogeneous model to investigate resource depletion, transitional regimes and intervention strategies.

In this first example, we will focus on a model of consumer with some shared resources that are critical for the survival of the consumer population. In this model, each consumer is characterized by their own value of parameter $c$, which determines the degree, to which the consumer depletes or restores shared resources. The model was initially proposed in (10) in the context of niche construction, and was later expanded in (11). It contains two coupled differential equations, written as follows:

$$
\underbrace{x_c(t)}_{\text{consumers}} \, ' = r \underbrace{x_c(t)}_{\substack{\text{population} \\ \text{growth rate}}} \, ( \underbrace{c}_{\text{consumption}} - \underbrace{\frac{N(t)}{kz(t)}}_{\substack{\text{dynamic} \\ \text{carrying} \\ \text{capacity}}} )
$$

$$
\underbrace{z(t)'}_{\substack{\text{shared} \\ \text{resource}}} = \underbrace{\gamma - \delta z(t)}_{\substack{\text{natural} \\ \text{resource turnover}}} + e \underbrace{\frac{N(t)(1-c)}{z(t)+N(t)}}_{\substack{\text{change in resource} \\ \text{caused by consumers} \\ \text{(depletion if c>1,} \\ \text{restoration if c<1)}}},
$$

$$(1.1)$$

where $N(t) = \sum_{\mathbb{A}} x_c(t)$ is the total population size over all possible values of parameter $c$. As one can see, it is assumed that the population grows according to the logistic growth function with a dynamic carrying capacity, determined by the shared resource $z(t)$. The consumers can either deplete the shared resource, or contribute to it, depending on the value of parameter $c$: $c > 1$ results in resource depletion, while $c < 1$ results in its restoration. The resource $z(t)$ also has a natural turnover rate, which can allow for sustainable coexistence of consumers with the resource. However, since increase in growth rate with respect to parameter $c$ creates an incentive for consumers to maximize resource consumption in the short term, this is likely to lead to destruction of shared resources, a notion that has been known as "the tragedy of the commons" (12–14). Situation when survival of the population depends on the over-depleted resource is known as "evolutionary suicide". It occurs when short-term increases in fitness due to resource overconsumption lead to eventual destruction of the shared resource and the population's extinction (15,16).

Several questions can be asked of this model, such as:

1) How will such a system behave depending on the number of over-consumers in it? What are the possible dynamical regimes that such a system can realize as it is heading for resource exhaustion and eventual population collapse?
2) Can we identify transitional regimes that can serve as warning signals of approaching collapse?
3) What, if any, intervention measures could be implemented to prevent the tragedy of the commons and possibly even evolutionary suicide?

Answering these questions requires a combination of both the classical approach of bifurcation analysis, and the ability to visualize evolutionary trajectories as the system evolves over time.

## Question 1. How will such a system behave depending on the number of over-consumers in it?

Answering this question can be achieved through conducting stability and bifurcation analysis, as has been done in (11). In this work, the authors progressively increased the value of parameter $c$ and observed a series of dynamical regimes, ranging from sustainable coexistence with the common resource with ever decreasing domain of attraction to sustained oscillatory regimes to collapse due to exhaustion of the common resource.

The results are summarized in Figure 1. In Domain 1, when the parameter of resource (over)consumption is small, the shared carrying capacity remains large, successfully supporting the entire population, since no individual is taking more resource than they replenish. In Domain 2, a parabolic sector appears near the origin, decreasing the domain of attraction of the non-trivial equilibrium point $A$. The population can still sustainably coexist with the resource even with moderate levels of over-consumption but the range of initial conditions, where it is possible, decreases. As the value of $c$ is further increased, the range of possible parameter values that allow sustainable coexistence with the common resource decreases and is now bounded by an unstable limit cycle, which appears around point $A$ through a catastrophic Hopf bifurcation in Domain 3, and via "generalized" Hopf bifurcation in Domain 6. Finally, in Domains 4 and 5, population extinction is inevitable due to extremely high over-consumption rates unsupportable by the resource.

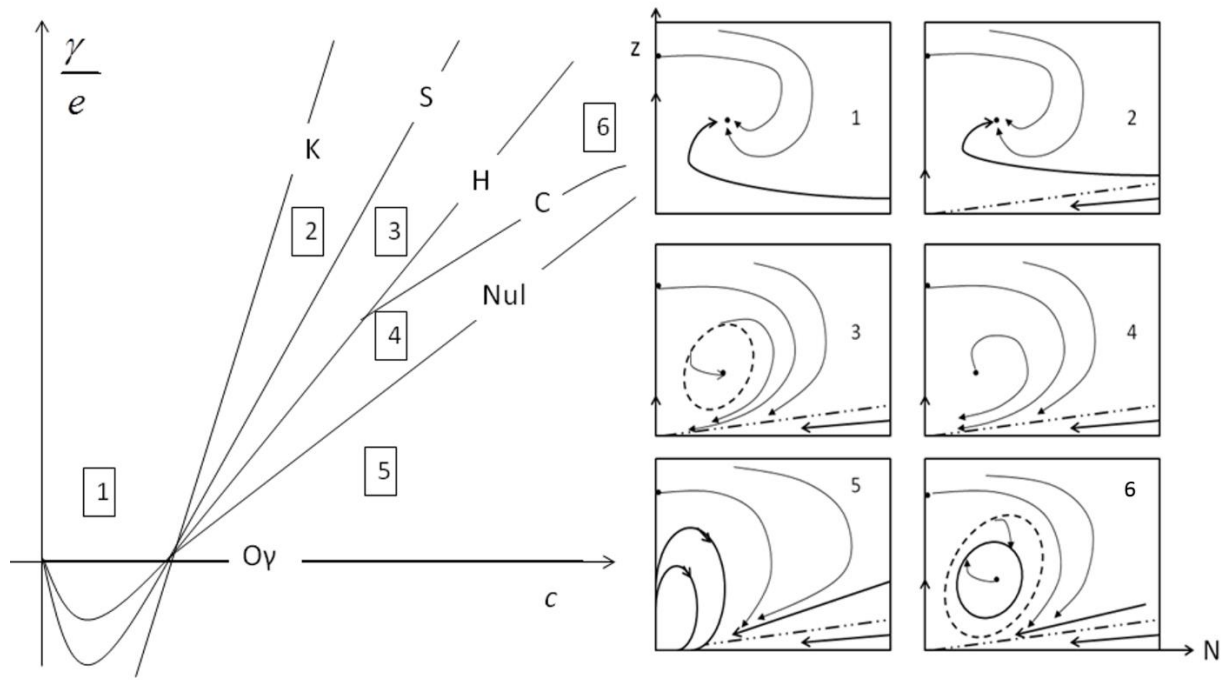

**Figure 1.** Bifurcation diagram of System (1.1) in the $(\gamma, c)$ and $(N, z)$ phase-parameter spaces for fixed positive parameters $e$ and $\delta$. The non-trivial equilibrium point $A$ is globally stable in Domain 1; it shares basins of attraction with equilibrium $O$ in Domains 2 and 3. The separatrix of $O$ and the unstable limit cycle that contains point $A$, serve, correspondingly, as the boundaries of the basins of attraction. Only equilibrium $O$ is globally stable in Domains 4, which also contains also unstable non-trivial $A$, and 5, which contains the elliptic sector. Domain 6 exists only for $\delta > 5 + \sqrt{24}$, where the stable limit cycle that is in turn contained inside an unstable limit cycle, shares basins of attraction with equilibrium $O$. Boundaries between Domains $K, S, H, Nul, C$ correspond respectively to appearance of an attractive sector in a neighborhood of $O$, appearance of unstable limit cycle containing $A$, change of stability of equilibrium $A$ via Hopf bifurcations, disappearance of positive $A$ and saddle-node bifurcation of limit cycles. The Figure is adapted from Figure 4 in (11).

## Question 2. Can we extrapolate transitional regimes that can serve as warning signals of approaching collapse?

Answering this question will require introducing population heterogeneity into the model to allow us to visualize evolutionary trajectories. As it stands, in the parametrically homogeneous case, analyzed in (11), the parameter value *c* is always a constant, and therefore the population will always remain in the corresponding domain of Figure 1.

However, let us introduce a keystone variable $q(t)$, such that $q(t)' = \dfrac{N(t)}{kz(t)}$. Then $x_c(t)' = rx_c(t)(c - q(t)')$. Consequently, $x_c(t) = x_c(0) = x_c(0)e^{ct - q(t)}$, and thus $N(t) = \int_{\mathbb{A}} x_c(t)dc = N_0 e^{-q(t)} \int_{\mathbb{A}} e^{ct} P_c(0)dc = N_0 e^{-q(t)} M_0(t)$, where $P_c(0) = \dfrac{x_c(0)}{N(0)}$ and $M_0(t) = \int_0^{\infty} e^{ct} P_c(0)$ is the moment generating function (mgf) of the initial distribution of clones within the population. The expected value of parameter *c* can then be calculated as $E^t[c] = \int_{\mathbb{A}} c P_c(t)dc = \int_{\mathbb{A}} c P_c(0) \dfrac{e^{tc}}{M_0(t)} dc = \dfrac{M_0(t)'}{M_0(t)}$.

The final system of equations thus becomes

$$N(t)' = N(E^t[c] - \frac{N(t)}{kz(t)})$$
$$z(t)' = \gamma - \delta z(t) + e\frac{N(t)(1 - E^t[c])}{z(t) + N(t)}, \tag{1.2}$$

where $E^t[c]$ is determined by the moment generating function of the initial distribution of clones in the population, as are consequently the dynamics of the entire system. Note that in comparison to the parametrically homogeneous System (1.1), in the parametrically heterogeneous System (1.2) the fixed value of the parameter c has been replaced by the expected value of *c* at each time instant *t*. It is easy to verify that the rate of change of $E^t[c]$ is equal to the variance of *c* at each time moment *t* in accordance to Fisher's fundamental theorem. Therefore, as the system evolves with time, the expected value of *c* will also change with each time step, causing it to "travel" through the phase-parametric portrait. A full analysis of this system, with all the derivations and proofs, was done in (11).

As an example, consider Figure 2, where the initial distribution for this model was taken to be truncated exponential, allowing for different maximal values of parameter *c*. The panels on the left depict the dynamics predicted by a parametrically homogeneous system, while the panels on the right depict the dynamics of a heterogeneous system.

Firstly, one can clearly observe the qualitative differences in predictions for population size and resource dynamics over time depending on the degree of population heterogeneity: a heterogeneous system survives longer, since it contains both over-consumers and under-consumers, with the latter delaying the collapse of the shared resource by "subsidizing" the over-consumers.

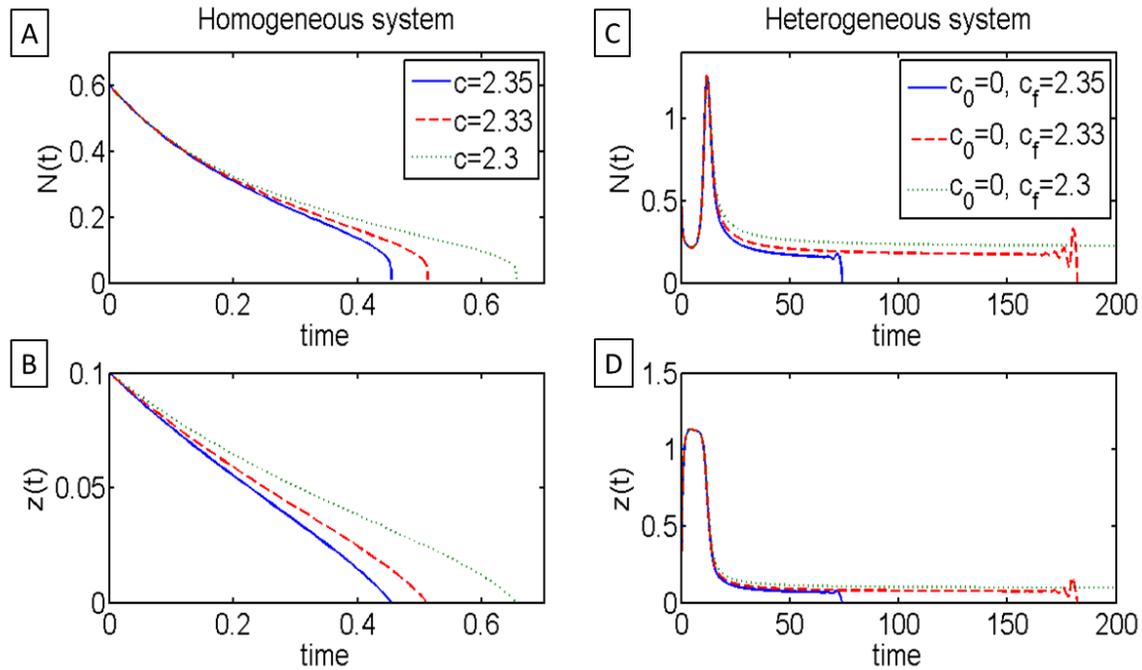

**Figure 2.** Comparison of trajectories of Systems (1.1) and (1.2) for different values of parameter $c$, or range of possible values of $c$, respectively. As one can see, while both parametrically homogeneous and heterogeneous populations can go extinct due to the exhaustion of common resource by over-consumers, time to extinction of a parametrically heterogeneous population is expected to be much larger. Moreover, in a parametrically heterogeneous system one can sometimes observe a transitional oscillatory regime preceding collapse, which is not observed in a parametrically homogeneous system. This figure is adapted from Figure 5 in (11).

Secondly, as one can see for the case, when the initial distribution of parameter $c \in [0 \; 2.33]$ (red dashed line), the system does in fact exhibit several transitional regimes as it goes through a period of growth through a period of seeming stability, to an oscillatory regime, which precedes population collapse. A closer look at this system in Figure 3 reveals that during this period of system stability, the expected value of parameter $c$ increases (Figure 3c), revealing the changes in population composition that will lead to its eventual collapse.

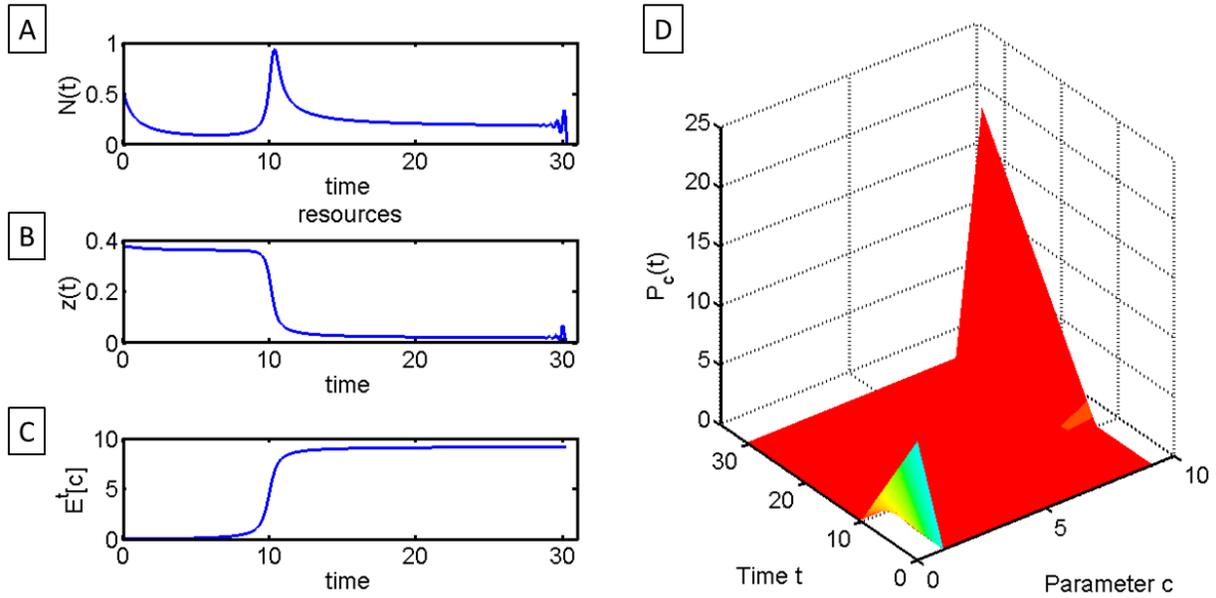

**Figure 3.** System (9) with initial truncated exponential distribution on the interval $c \in [0 \; 2.5]$. An example of what transitional regimes the System can go through before the population crashes, depicting (A) trajectories for the total population size *N(t)*, (B) total amount of renewable resource *z(t)*, (C) expected value of the parameter c, and (D) the change over time in distribution of various clone types within the population. Initial conditions fall within the parameter range of Domain 6 of the phase-parameter portrait of the non-distributed system (Figure 1). Since the rate of natural resource decay is high, it takes more time even for the most efficient consumer to ''get to it'', and so the population survives longer, and the transitional regimes are more evident. This Figure is adapted from Figure 7 in (11).

It is not always clear that system collapse is approaching, and so one has to learn to recognize early warning signals, such as increased flickering and data auto-correlation (17–19), in order to try and prevent the tragedy of the commons. Application of the HKV method to relevant systems of ODEs allows to visualize exactly how the system passes through these dynamical regimes as it evolves. One can see that while changes in population size and the resource over time may seem to give no cause for alarm, the mean value of the parameter of over-consumption may signal trouble: the system will be recalibrating towards maximizing c, and as soon as the buffer capacity of the resource (in this case it is proportional to natural resource restoration and decay rates) is exhausted, both the population and the resource collapse.

Notably, in (19), Dakos et al. analyzed eight ancient abrupt climate shifts and showed that in each case they were preceded by a characteristic slowing down of fluctuations preceding the actual shift, similarly to behaviors predicted in Figure 2, suggesting that even a relatively simple parametrically heterogeneous model can provide meaningful results and even qualitative, if not quantitative, predictions.

## Question 3. What, if any, intervention measures could be implemented to prevent the tragedy of the commons and possibly even evolutionary suicide?

In order to address this question, we can introduce a punishment/reward function that can affect individuals in the population based on the value of parameter of over-consumption $c$. The updated system of equations would look as follows:

$$x_c(t)' = \underbrace{r}_{\substack{\text{population} \\ \text{growth rate}}} \underbrace{x_c(t)}_{\text{consumers}} (\underbrace{c}_{\text{consumption}} - \underbrace{\frac{N(t)}{kz(t)}}_{\substack{\text{dynamic} \\ \text{carrying} \\ \text{capacity}}}) + \underbrace{x_c(t)f(c)}_{\text{punishment/reward}}$$

$$z(t)' = \underbrace{\gamma - \delta z(t)}_{\substack{\text{natural} \\ \text{resource turnover}}} + \underbrace{e\frac{N(t)(1-c)}{z(t) + N(t)}}_{\substack{\text{change in resource} \\ \text{caused by consumers} \\ \text{(depletion if c>1,} \\ \text{restoration if c<1)}}}.$$

$$\underbrace{\phantom{z(t)}}_{\substack{\text{shared} \\ \text{resource}}}$$

$$(1.3)$$

This way, depending on the form of the punishment function $f(c)$, one can try to impose punishment on over-consumers, reward under-consumers, and hopefully be able to maintain the population in a range where it can sustainably co-exist with its dynamic resource.

In (20), the authors investigated three types of punishment/reward functions:

1) Moderate punishment $f(c) = a\dfrac{1-c}{1+c}$

2) Severe punishment/generous reward $f(c) = a(1-c)^3$, where the parameter $a$ denotes the severity of implementation of punishment on individuals with the corresponding value of parameter $c$

3) Separating punishment and reward: $f(c) = \rho(1-c^\eta)$. This functional form allows to separate the influence of reward for under-consumption, primarily accounted for with parameter $\rho$, and punishment for over-consumption, accounted for with parameter $\eta$.

We evaluated the effectiveness of these three types of punishment/reward functions on system evolution and calculated predicted outcomes for different initial distributions of clones within the population, which were taken to be truncated exponential and Beta distributions. The initial distributions were chosen in such a way as to give significantly different shapes of the initial probability density function; in applications they should be matched to real data, when it is available. We observed that the intensity of implementation of punishment/reward has to differ for different initial distributions if one is to successfully stop over-consumption, and so in order to be able to make any reasonable predictions one needs to understand what the initial composition of the affected population is (see Figure 4).

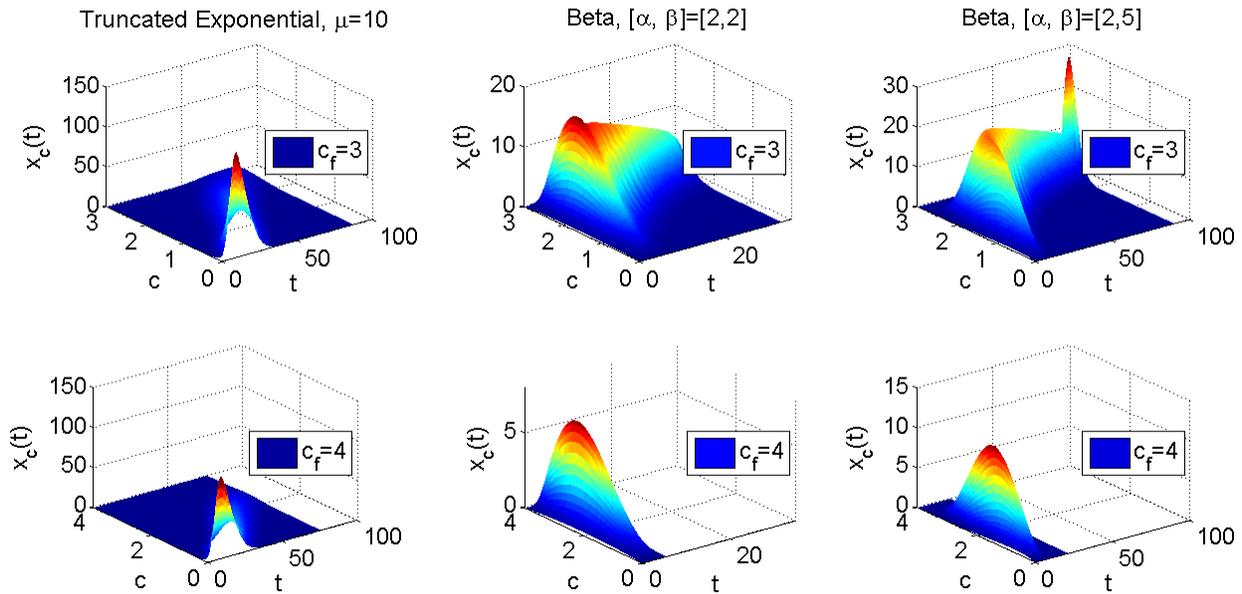

**Figure 4.** The importance of evaluating the range of possible values of $c_{f}$, illustrated for different initial distribution. Punishment function is of the type $f(c) = \rho(1 - c^{\eta})$, where $\rho = 0.6$, $\eta = 1.2$. Initial distributions are taken to be truncated exponential with parameter $\mu = 10$, and Beta with parameters $\alpha = 2, \beta = 2$ and $\alpha = 2, \beta = 5$; $\rho = 0.6$, $\eta = 1.2$. The top row corresponds to $c \in [0,3]$; the bottom row corresponds to $c \in [0,4]$. Figure adapted from Figure 12 in (20). The simulations were conducted by Benjamin Morin.

We also observed that severe punishment/generous reward approach was much more effective in preventing the tragedy of the commons than the moderate punishment/reward function, particularly for the cases, when over-consumers were present at higher frequencies (such as both Beta initial distributions). This comes not only from the severity of punishment but also from the fact that moderate punishment allows more time for the over-consumers to replicate, and thus by the time the punishment has an appreciable effect, the population composition had changed, and the moderate punishment will no longer be effective. So, in punishment implementation one needs to take into account not only the severity of punishment but also the time window that moderate punishment may provide, allowing over-consumers to proliferate. Within the frameworks of the proposed model, moderate implementation of more severe punishment/reward system is more effective than severe implementation of moderate punishment/reward. Complete analysis of System (1.3) and further simulations are reported in detail in (20).

In (20), we proposed just one way to try and modify individuals' payoffs in order to prevent resource over-consumption - through inflicting punishment and reward that affects the growth rates of clones directly. This approach can be modified depending on different situations, inflicting punishment or reward based not just on the intrinsic value of $c$ but on total resource currently available.

To summarize, a system of two equations describing the dynamics of consumers depleting and replenishing shared resources was simple enough to allow complete analysis and generation of a comprehensive bifurcation diagram. However, application of HKV-method allowed to qualitatively expand the realm of questions that the model could answer, which could have significant practical applications, particularly in the area of sustainability.

## Example 2. Mixed strategies and natural selection

In this example, we will explore a reformulation of the model of consumer-resource interactions within the context of strategy selection. Specifically, we will look at a model that deals with the question of strategies of resource allocation.

Broadly speaking, the two "main" strategies that can be adopted by different species in response to different selective pressures that come from their environment are either to invest the resources into rapid proliferation, which has been suggested to be the preferable strategy in unstable environments, or into physiological maintenance and increasing environmental carrying capacity at the expense of rapid proliferation, which would allow maximizing fitness in more stable conditions (21–23). The main criticism of this theory came from empirical studies:

however intuitive the heuristic may seem, the adaptations that were predicted by either selective strategy were rarely if ever observed in nature (24). Nevertheless, there may be merit to this theory if one focuses not on looking for pure strategies but rather explores a continuum. In (25), we described such a situation by introducing parameter $\alpha$ to denote the strategy of investing available resource solely in reproduction, and $(1-\alpha)$ to denote the strategy of investing the available resource primarily into increasing and maintaining the physiological carrying capacity. We considered the dynamics over time of a population of individuals $x_\alpha$ characterized by their particular value of "strategy" $\alpha$, which can fall anywhere within the continuum $\alpha \in [0,1]$. When $\alpha = 0$, each individual was assumed to grow according to the functional form $r(c_2 \frac{z}{N+z} - \phi)$, where $N(t) = \int_A x_\alpha d\alpha$ is the total population size of all individuals, and $z(t)$ is a shared resource. As one can see, in this case shared resources $z(t)$ are allocated to increasing the rate of proliferation of individuals $x_\alpha$. When $\alpha = 1$, each individual grows according to the logistic growth function with dynamic carrying capacity, given by $r(c_1 - \frac{bN}{kz})$. If the individual uses both strategies with probabilities $\alpha$ and $(1-\alpha)$ respectively, i.e., uses some of the resource towards rapid proliferation and some towards physiological maintenance, then the per capita growth rate of each $\alpha$-clone is given by $\alpha r(c_1 - \frac{bN}{kz}) + (1-\alpha)(c_2 \frac{z}{N+z} - \phi)$.

Shared resource $z(t)$ is assumed to have a natural turnover rate, being replenished at some constant rate $\gamma$ and decaying at a rate $\delta z(t)$, as well as be consumed or restored by all the individuals. The consumption-restoration process is accounted for by the term $e \frac{N(t)(1-c)}{z(t)+N(t)}$; as the number of consumers increases, the amount of resource will increase or decrease depending on the value of parameter $c_1$ for $\alpha$-strategy, or $c_2$ for $(1-\alpha)$-strategy. Full derivation of the system is given in (25).

The final model then becomes

$$\underbrace{\frac{dN(t)}{dt}}_{\substack{\text{population}\\\text{size}}} = rN(t)(\ \underbrace{\alpha r(c_1 - \frac{bN}{kz})}_{\substack{\text{proportion of individuals}\\\text{investing resource directly}\\\text{in proliferation}}} + \underbrace{(1-\alpha)(c_2 \frac{z}{N+z} - \phi)}_{\substack{\text{proportion of individuals investing}\\\text{resource in physiological maintenance}}}\ ),$$

$$\underbrace{\frac{dz(t)}{dt}}_{\substack{\text{shared}\\\text{resource}}} = \underbrace{\gamma - \delta z(t)}_{\substack{\text{natural resource}\\\text{turnover}}} + eN(t)(\ \underbrace{\frac{\alpha(1-c_1)}{N(t)+z(t)}}_{\substack{\text{resource consumed/restored}\\\text{by individuals investing it in}\\\text{proliferation}}} + \underbrace{\frac{(1-\alpha)(1-c_2)}{N(t)+z(t)}}_{\substack{\text{resource consumed/restored}\\\text{by individuals investing it in}\\\text{physiological maintenance}}}\ ).$$

(1.4)

Several questions can now be asked of such a model, such as:

1) If one allows for the possibility of resource overconsumption, which strategy is preferable for avoiding population collapse and consequently the tragedy of the commons?

2) Which strategy (allocating shared resources towards rapid proliferation, or towards slower proliferation but increased physiological and environmental maintenance) is more likely to become dominant as a result of natural selection?

Similarly to the previous example, answering these questions will require the use both of classical analytical methods and the HKV method.

## Question 1. If one allows for the possibility of resource overconsumption, which strategy is preferable for avoiding population collapse as a result of the tragedy of the commons?

Answering this question, as in the previous case, can be achieved through conducting stability analysis and in particular by evaluating how system behavior changes with regards to changes in strategy parameter $\alpha$ and concurrent changes in parameters of resource consumption $c_1$ and $c_2$. The obtained bifurcation diagram (see Figure 5) describes the possible dynamical regimes of a population that is homogeneous with respect to $\alpha$.

An important conclusion from the bifurcation analysis is that the main qualitative regimes of behaviors and also the sequence in which they appear as the parameters of (over-) consumption change are very similar for both extreme cases. However, wider domains of sustainable coexistence with shared resource were identified for the second strategy of allocating the resources towards physiological maintenance even under increasing values of parameters of resource (over)consumption. This suggests that at least in the case of a parametrically homogeneous system, investing in physiological maintenance might be a more sustainable strategy.

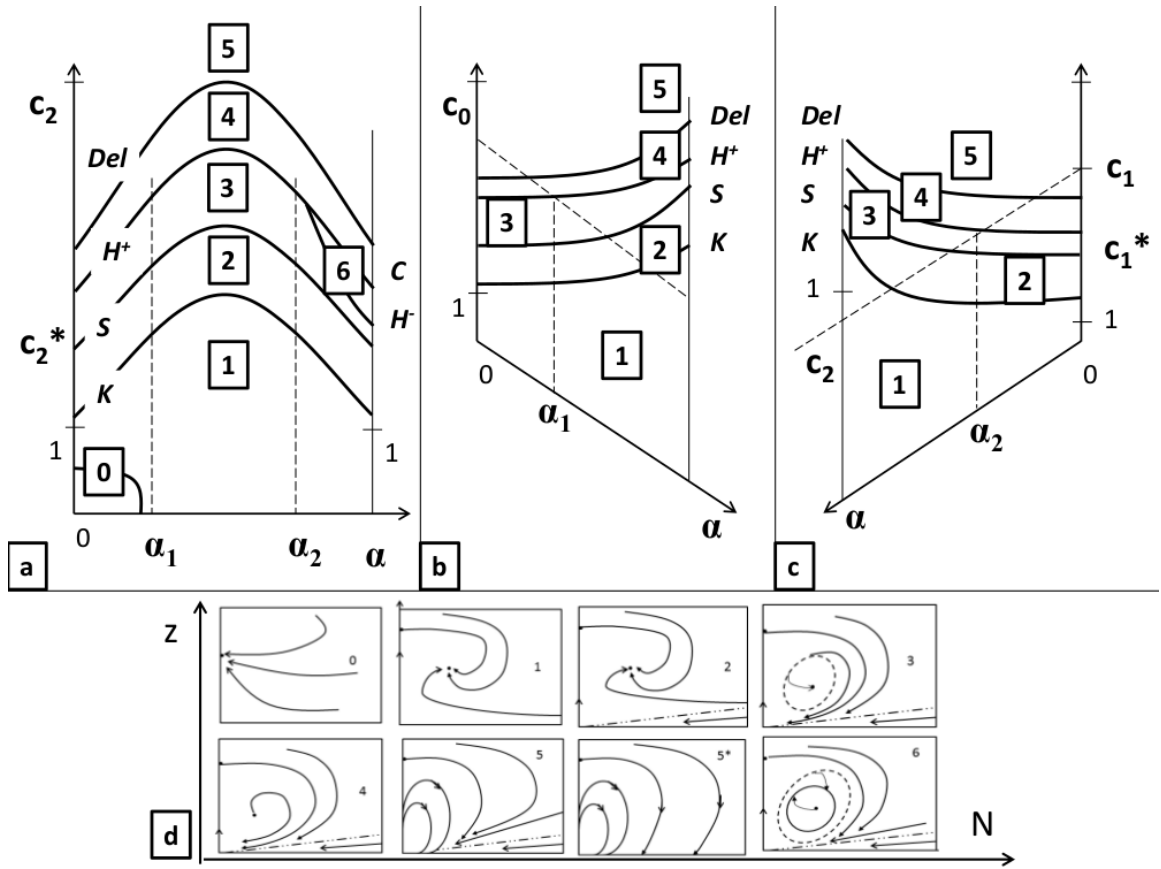

**Figure 5.** Bifurcation diagram of the System (1.4). (a),(b),(c) present schematically $(c_1, c_2, \alpha)$ parameter portraits for fixed values of $\gamma, \delta, e = 1$ and (d) represents the corresponding typical phase portraits. In Domain 1 there exists a non-trivial globally attracting equilibrium point $A_\alpha$. Domains 2 and 6 are the regions of bistability; in Domain 2, there is a nontrivial stable node, while in Domain 6 there exists a stable oscillatory regime. In these regions population survival is conditional on the initial population size and the initial amount of resource. In Domain 3, an unstable limit cycle is formed around the point $A_\alpha$, shrinking the range of possible initial conditions that will lead to sustainable population survival. In Domain 4, point $A_\alpha$ is unstable, so any perturbation will lead to population collapse. In Domain 5, an elliptic sector appears, which implies that a population is bound for extinction regardless of initial conditions. Finally, Domain 0 corresponds to the case, when only trivial equilibrium $B(0, \frac{\gamma}{\delta})$ is globally attractive, which is of no biological interest.

Question 2. Which strategy (allocating shared resources towards rapid proliferation, or towards slower proliferation but increased physiological and environmental maintenance) is more likely to become dominant as a result of natural selection?

The answer to this question required application of the HKV-method to distribute parameter $\alpha$ (the details of this relatively complex transformation are given in (25)). The resulting parametrically heterogeneous system allowed exploring the changes in predicted evolutionary trajectories depending on the initial composition of the population with respect to different strategies.

The results of these simulations revealed that in this system, the direction of population evolution is extremely sensitive to initial population composition (see Figure 6). This suggests that even though one strategy might be preferable for a parametrically homogeneous population, in a parametrically heterogeneous case the direction of the evolutionary trajectory is determined primarily by population composition, i.e., by initial distribution of clones with the population. This can be interpreted as "founder effect", when the initial composition of the small population determines the subsequent evolutionary trajectory of the population over time (26).

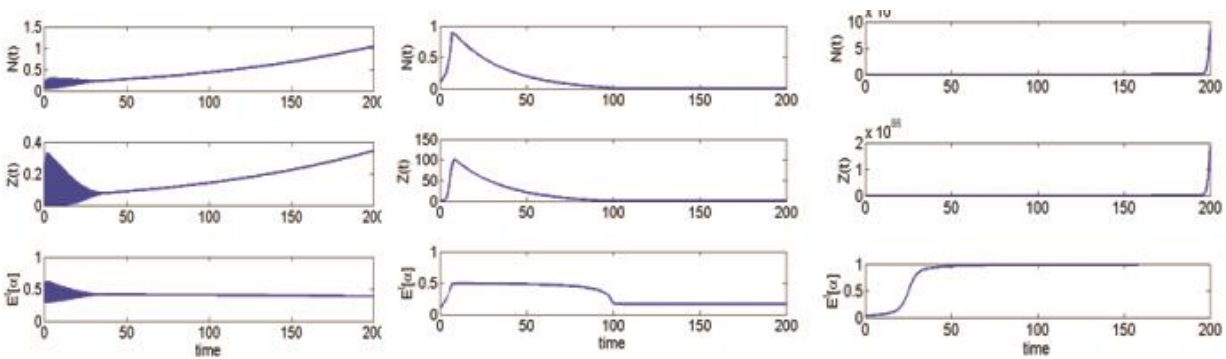

**Figure 6.** The effects of difference in the initial composition of the population with respect to different strategies. Different initial distributions were chosen to be (a) uniform initial distribution (b) truncated exponential initial distribution, with parameter $\mu = 1.1$ (note: population crashes after time t = 32) and (c) truncated exponential initial distribution, with parameter $\mu = 10.1$. Initial conditions are such as to fall within Domain 1. All parameters held constant at r = 1, e = 1, b = 1, k = 1, $N_0 = .1$, $c_2 = .2$, $c_1 = .6$, d = 1, p = 1, $\phi = 0.14$. One can see that the initial composition of the population can have dramatic effects on the direction in which the population will evolve over time. (Note: the values of μ were chosen arbitrarily for illustrative purposes). This Figure is adapted from Figure 6 in (25).

# Example 3. Oncolytic viruses

In this final example, we will look at a model presented in (27), which describes the dynamics of cancer cells that can be infected by an oncolytic virus, i.e. a virus that can specifically infect and kill cancer cells but leave normal cells unharmed (28–30). The proposed model considers two types of cells, infected and uninfected cancer cells, growing in a logistic fashion. The system is described by the following two equations:

$$\underbrace{\frac{\mathrm{dX}}{dt}}_{\substack{\text{uninfected} \\ \text{cancer cells}}} = \underbrace{r_1 X(1 - \frac{X+Y}{K})}_{\substack{\text{logistic growth to shared} \\ \text{carrying capacity K}}} - \underbrace{\frac{\beta XY}{X+Y}}_{\substack{\text{rate of virus} \\ \text{transmission}}}$$

$$\underbrace{\frac{dY}{dt}}_{\substack{\text{infected} \\ \text{cancer cells}}} = \underbrace{r_2 Y(1 - \frac{X+Y}{K})}_{\substack{\text{logistic growth to shared} \\ \text{carrying capacity K}}} + \underbrace{\frac{\beta XY}{X+Y}}_{\substack{\text{rate of virus} \\ \text{transmission}}} - \underbrace{\delta Y}_{\substack{\text{death of} \\ \text{infected} \\ \text{cancer cells}}}$$

(1.5)

where $X$ is the size of the uninfected cancer cell population; $Y$ is the size of the infected cancer cell population; $r_1$ and $r_2$ are the maximum per capita growth rates of uninfected and infected cells, respectively; $K$ is the carrying capacity; $\beta$ is the transmission coefficient, which may also include the replication rate of the virus; and $\delta$ is the rate of additional infected cell death rate as caused by the virus.

The following questions can be asked and answered by this model:

1) What are the transitional regimes that occur as the cancer cell population gains resistance to the virus? Can we use the model to infer dynamics of evolution of resistance?
2) Why are cytotoxic therapies effective in some patients and not others?

Similarly to the previous cases, both approaches – classic bifurcation analysis and modeling of heterogeneity – will be necessary to answer these questions.

**Question 1. What are the transitional regimes that occur as the cancer cell population gains resistance to the virus? Can we use the model to infer dynamics of evolution of resistance?**

In order to answer this question, a bifurcation analysis needs to be performed. A full bifurcation diagram can give a sense of what transitional regimes a population goes through as it moves from the area of phase-parameter space of tumor elimination to that of tumor growth, similarly to the previous examples.

The complete phase-parameter portrait of System (1.5) is shown in Figure 7. The model exhibits all possible outcomes of life cycle of infected and uninfected cells. In Domains 1 and 2, there is no effect of the viral infection on the tumor; in Domains IV and V, tumor load is stabilized and even reduced. Complete elimination of the tumor can be observed in domain VIII. Furthermore, there are two domains (domains III and VII) where the final outcome crucially depends on the initial conditions and can result either in failure of virus therapy or in stabilization (domain III) and elimination (domain VII) of the tumor.

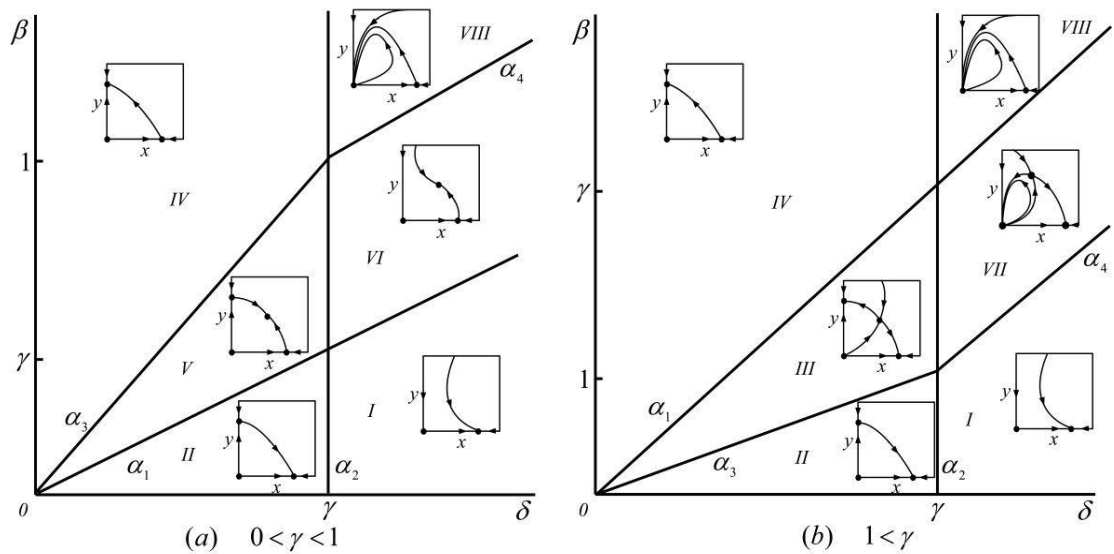

**Figure 7.** Bifurcation diagram of the parametrically homogeneous system reported in (27) and reproduced here in System (1.5). All possible outcomes of oncolytic virus infection are as follows: no effect on the tumor (domains I and II), stabilization or reduction of the tumor load (domains IV and V), and complete elimination of the tumor (domain VIII). Moreover there are two domains (domains III and VII) where the final outcome crucially depends on the initial conditions and can result either in failure of virus therapy or in stabilization (domain III) and elimination (domain VII) of the tumor. Figure is adapted from Figure 1 in (27).

Introduction of population heterogeneity with respect to parameter of viral transmission $\beta$ allowed more complete visualization of possible evolutionary trajectories of the tumor. For example, in simulations in Figure 8, parameter values were chosen in such a way as to start in domain VIII, where complete tumor elimination occurs. However, as the system evolved, the dynamics crossed from the domain of complete tumor elimination (VIII) to that of bistability (domain VII) to end up in the domain of tumor escape (domain I). Furthermore, differences in variances of initial distributions resulted in changes in predicted tumor dynamics, with lower variances corresponding to longer periods of near-negligible tumor sizes, a dynamical regime that can be interpreted as tumor dormancy, or "cancer without disease", when a tumor is present in the tissue but it not growing (31–33).

More broadly, we can infer from the bifurcation analysis and subsequent simulations that as the tumor population becomes resistant, it travels through the various domains described in Figure 7, allowing us to better understand the transitional regimes of evolution of resistance.

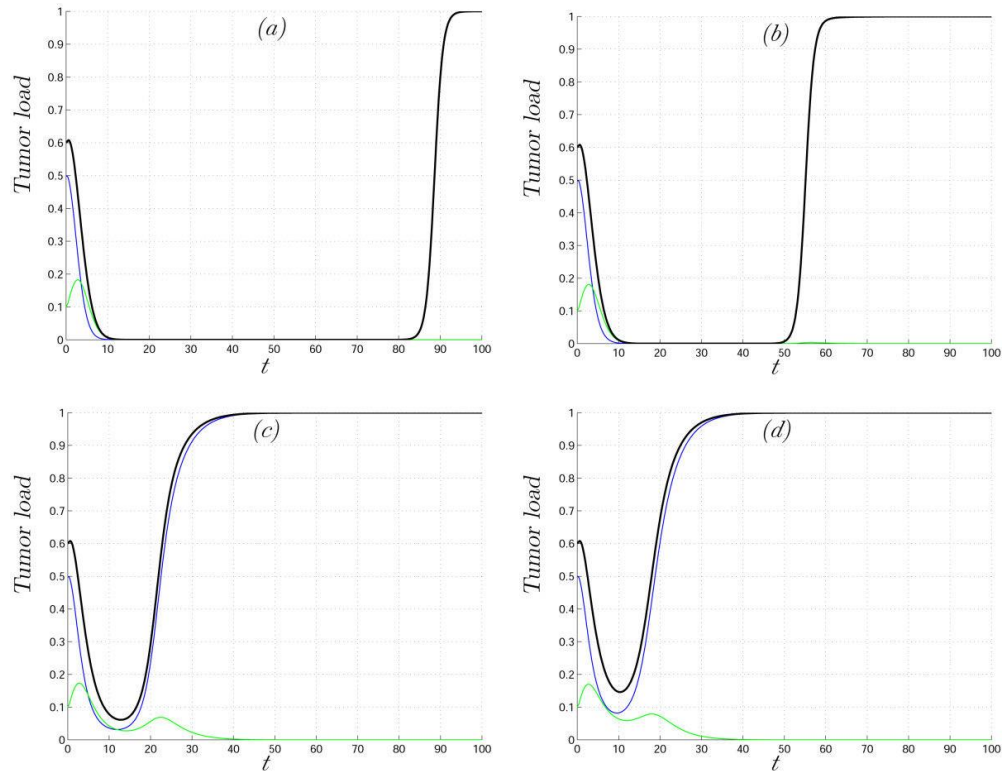

**Figure 8.** Solutions of parametrically heterogeneous system reported in (27) with Gamma-distributed parameter of transmission of the oncolytic virus. The solutions here reflect the fact that the degree of heterogeneity plays an important role in the model dynamics. The parameter values and initial conditions are the same for all four simulations; the difference comes from different initial variances of the initial distribution; the greater the initial variance the faster we reach the unfavorable domain I. The figure is adapted from Figure 2 of (27).

## Question 2. Why are cytotoxic therapies effective in some patients and not others?

The answer to this question came from further simulations conducted by the authors, where they showed that *initial composition* of the population may be one of the culprits underlying emergence of resistance in some tumors but not others. Specifically, in Figure 9 they showed that *differences in variance of the initial distribution* of cell clones within the population can lead to qualitatively different final outcomes of oncolytic therapy. These results may shed some light on the question of variability in therapeutic successes for other interventions, a topic that is of vital importance.

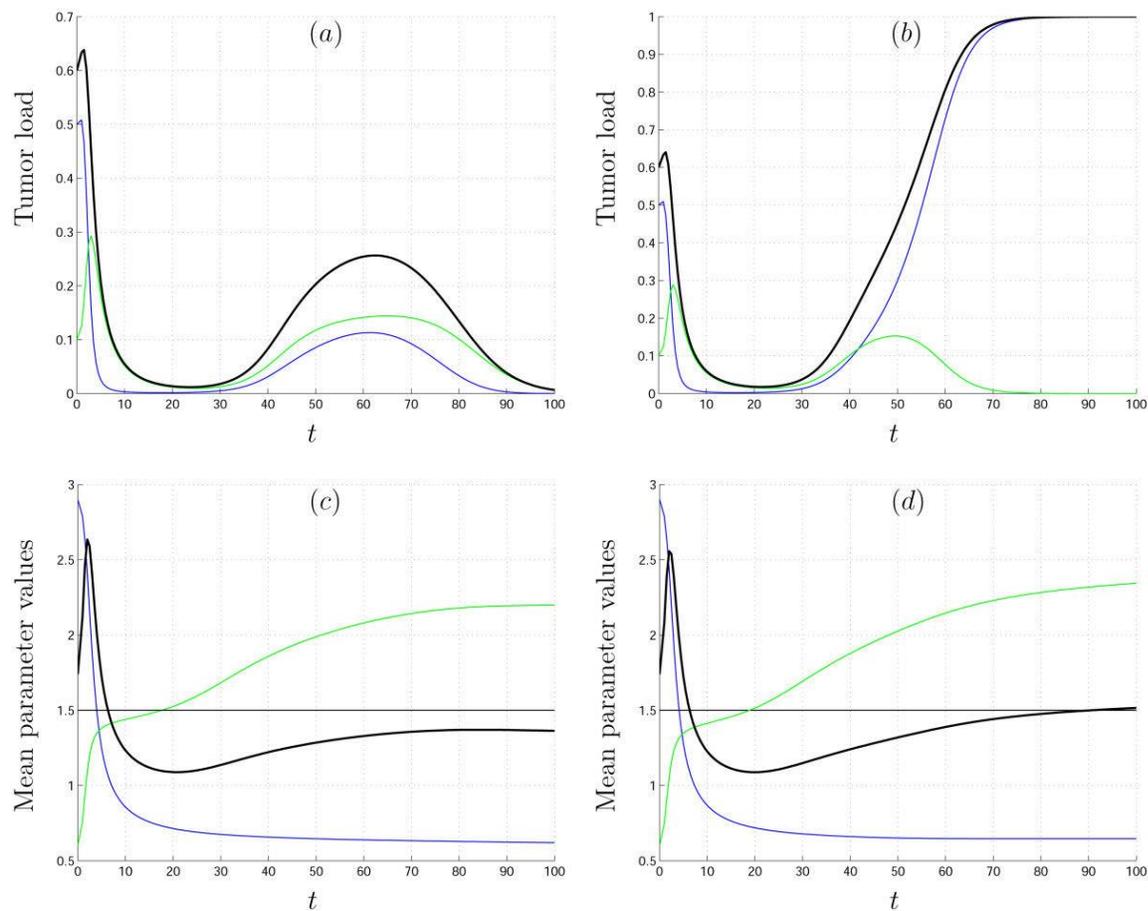

**Figure 9.** Solutions of parametrically heterogeneous system presented in (27) with both uninfected cell specific and infected cell specific distributions of transmission coefficient. The initial conditions and parameter values are the same for both cases; the two cases differ only in the initial variance of the initial distribution of the transmission coefficient. As one can see, even a small difference in the variance of the initial distribution of the cell clones may yield dramatically different results: in the first case, the tumor is cured, whereas, in the second case, virus therapy fails. The figure is adapted from Figure 7 in (27).

# Conclusions

Classic techniques for analysis of dynamical systems can provide critical insights into the possible dynamical regimes that a system can realize. Unfortunately, doing full bifurcation analysis is labor intensive and is not always possible due to increasing complexities of many models. However, there already exists a very rich body of literature of fully analyzed parametrically homogeneous models in many fields, including ecology (34,35), epidemiology (36–38), among others. As the examples presented here demonstrate, even relatively simple two-dimensional systems can reveal rich, unexpected and meaningful behaviors. Application of the HKV-method to introduce population heterogeneity in a meaningful way, and utilizing previously performed analysis can reveal a new layer of understanding of many existing models that was not accessible before. This of course is possible only if we ask the right questions.

# Acknowledgements

This research received no external funding.